# AUDIT MATURITY MODEL

Bhattacharya Uttam, Rahut Amit Kumar, De Sujoy

Cognizant Technology Solutions, Kolkata, India
uttam.bhattacharya@cognizant.com / amit.rahut@cognizant.com / sujoy.de@cognizant.com

## ABSTRACT

*Today it is crucial for organizations to pay even greater attention on quality management as the importance of this function in achieving ultimate business objectives is increasingly becoming clearer. Importance of the Quality Management (QM) Function in achieving basic need by ensuring compliance with Capability Maturity Model Integrated (CMMI) / International Organization for Standardization (ISO) is a basic demand from business nowadays. However, QM Function and its processes need to be made much more mature to prevent delivery outages and to achieve business excellence through their review and auditing capability. Many organizations now face challenges in determining the maturity of the QM group along with the service offered by them and the right way to elevate the maturity of the same. The objective of this whitepaper is to propose a new model –the Audit Maturity Model (AMM) which will provide organizations with a measure of their maturity in quality management in the perspective of auditing, along with recommendations for preventing delivery outage, and identifying risk to achieve business excellence. This will enable organizations to assess QM maturity higher than basic hygiene and will also help them to identify gaps and to take corrective actions for achieving higher maturity levels. Hence the objective is to envisage a new auditing model as a part of organisation quality management function which can be a guide for them to achieve higher level of maturity and ultimately help to achieve delivery and business excellence.*

## 1. INTRODUCTION

For any world class organization, quality compliance to its standard software process [1] is considered as a basic hygiene factor. ISO [2] and CMMI [3] are official certification/assessment for this which each business unit must ensure.

In today's business scenario, focus of the Quality Assurance (QA) function needs to be elevated from traditional compliance related aspects to more value added services to justify its presence to meet business objectives. Audit function, instead of ensuring mere compliance needs to be much more matured to prevent delivery outage and to achieve business excellence which are the call of the day for survival and to prove oneself best in class in the industry.

To keep the quality function as one of the essential business functions, the focus of Quality Assurance activities (audit, review etc.) should be elevated towards higher quality of deliverables and higher performance by strengthening process maturity and quality of data. That way, prevention of delivery outage can be achieved through proactive identification of the risks associated with delivery management, product quality and process adherence. Furthermore, focusing on business excellence by business risk assessment along with management of client's expectation will help in reaching highest maturity.

## 2. AUDIT MATURITY MODEL (AMM)

Audit Maturity Model (AMM) framework will provide organizations with an assessment of the maturity of audit and review processes / capabilities in the perspective of auditing capability, along with recommendations for achieving higher levels of maturity. This will ensure assessment of not only the basic hygiene factors but also of engagement maturity and business excellence.

At the bottom level, audit / review activities are informal, chaotic and adhoc. Reviews and audits are carried out mainly on reactive basis to understand and correct burning project issues. Hence success of the reviews and audits depends on the skill of the people conducting the reviews & audits. There is no Software Quality Assurance (SQA) group defined to assess the audit process. This level can be called as Level 1 initial. There is no formal auditing team to meet the basic objective.

At level 2, localized standards of reviews and audits have been recognized, best practices for different reviews and audits are identified and software quality assurance group formed to make it more manageable. At this level, reviews and the audit activities are much more disciplined than level 1 and meet all basic need by focusing on setting up of a standard / compliant process. At this level, SQA Team exists and the objective of audits is to ensure verbatim compliance to meet all basic hygiene. This type of audit can be called as Disciplined Audit, and are carried out by members of the SQA group.

At the next level, the audit activities are completely standardized and consistent. Reviews and audits are now much more compliant to many international standards. The audit function now focuses on process maturity through repeatable results and increasing scope of audits. Sets of well-defined and documented standard processes are established and the auditing activities are now formal. The main objective of audits at this level is to ensure process maturity, and audits are carried out by experienced members of the SQA group.

Level 4 is much more matured and now the focus of audits shifts to proactive risk identification to ensure product quality and maturity. Delivery management with stable product quality and process adherences are key aspects to prevent delivery outage at this level. Audits here are carried out by senior members of the SQA team along with seasoned project and delivery managers.

At the level 5, there is a paradigm shift audits focus on business excellence rather than process maturity or delivery maturity. Assessment of business risks in the area of Finance, Customer Relations, Employee, Infrastructure, and Security are the main objective at this level. At this level, audits are carried out by senior management team members.

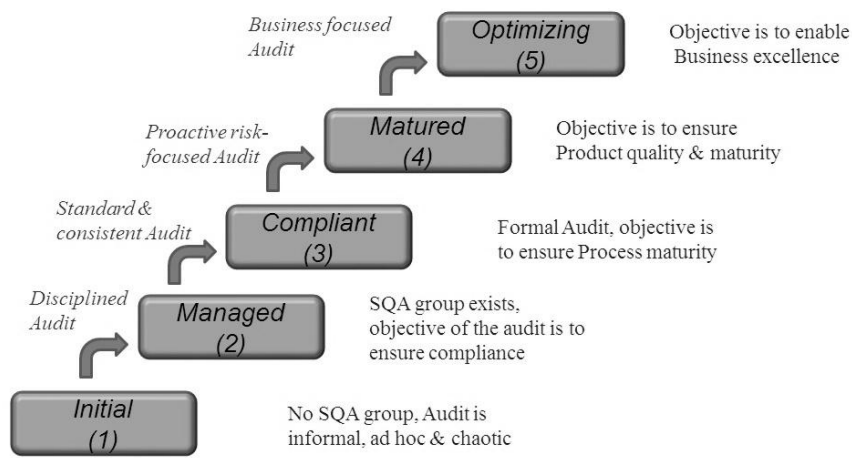

Fig. 1. Audit Maturity Model (AMM)

## 3. CHARACTERISTICS OF THE AUDIT MATURITY MODEL (AMM)

In Audit Maturity Model, lower levels of maturity form the basis of a higher maturity level. Hence, it is not possible to achieve higher maturity level if a lower level is skipped. Hence assessment of reviews / audit maturity can be achieved stage wise from level 2 to upwards. Followings are few characteristics of Audit Maturity Model:

- This audit model automatically helps to ensures process compliance. Organizations assessed at CMMI level 2 or certified in ISO, AMM helps to ensure compliance to the organization standard software process, thereby confirming basic hygiene.

- At lower maturity level, basic risks are identified and mitigation actions are planned so that the higher maturity level can focus on more vital aspects and identify more business-critical risks.

- Delivery management, product quality and process adherences risks are proactively identified till maturity level 4 which help in enhancing execution maturity.

- Maturity Level 5 reinforces client expectations by identifying and mitigating business risks in the area of Finance, Customer Relations, Employee, Infrastructure, and Security.

## 4. IMPLEMENTATION APPROACH OF AUDIT MATURITY MODEL (AMM)

The assessment of maturity reviews / audit activities is an examination of different goals defined at different levels by a trained team of professionals using Audit Maturity Model framework as a basis for determining strengths and weaknesses of an organization. This will help to identify gaps at different levels in the framework. Weaknesses can be analyzed and proper action items can be implemented to close the gaps and thus achieve maturity of a particular level, as also proceed to higher maturity levels.

The relationship between the different audits to be conducted and focus area of Audit Maturity Model (AMM) is demonstrated in the figure below. At the bottom of sharp end of V, there is no formal audit or risk assessment. At the next level, the audit is called Discipline Audit to check compliance of level 2 goals of focusing on process compliance and data quality. This can be done through desktop audit by auditing, collecting and analyzing the data for projects of the organisation. In a mature organization, this can also be performed remotely by extracting necessary data from defined tools. The risk of non-compliance of process and data quality needs to be shared with the corresponding stakeholders to identify and implement further corrective and preventive actions.

| Focus Area | | Audit Name | |
|---|---|---|---|
| Level 5: Focusing on Finance, Customer Relations, Employee, Infrastructure, Security | Business Risk Assessment | | Engagement Maturity Audit |
| Level 4: Preventing delivery outage | Delivery Management Risk Assessment | | Execution Maturity Audit |
| Level 3: Focusing on quality of deliverables | Product Quality Risk Assessment | | Process, Work Product & Delivery Audits |
| Level 2: Focusing on Process Maturity and Data Quality | Process Compliance Risk Assessment | | Desktop Audit |
| Level 1: Reactive audit, adhoc and chaotic | Adhoc/ reactive audit/ no Risk Assessment | | |

Fig. 2. Implementation Approach of Audit Maturity Model (AMM)

At the next level, different types of audit are executed like Process Audit which focuses on process maturity, Work Product Audit which ensures quality of all deliverables; and finally Delivery Audit which controls quality of the delivered product or services. These standard and consistent audits can focus on quality of deliverables with process maturity by identifying risks of product quality.

Once the focus has shifted completely from process compliance to process maturity, and quality of deliverables are assured by level 2 and level 3 audit capability of AMM implementation, audits now need to focus on product quality and maturity by identifying proactive risks of delivery management. This Execution Maturity Audit includes product quality with delivery management aspects to prevent delivery outage.

At the highest level, the objective is to identify and assess business risks associated with financial performance, the relationship between various groups in the program / project, customer relationship, staffing, infrastructure, business continuity and security, etc. through Engagement Maturity Audit. At this level, execution maturity transforms to engagement maturity so as to achieve business excellence. The Quality Assurance function aided by senior management must also work proactively at this stage to align the vendors / suppliers, the organization and its customers.

The audit function must identify the aforesaid risks proactively and escalate through defined path to the stakeholders in coordination with project senior team members. The risks must be identified and mitigated proactively before they affect the business or customer. Detailed audit checklists can be made based on different goals and these can be used to dig to a granular level to make the audits more stringent. The appraisal process also needs to be mature enough to produce consistent results through these audits for elevating themselves to the next level.

When planning an audit of the AMM framework, the scope of the disciplines to be included needs to be determined. Other considerations include whether the audit team will consist of members internal or external to the organization; individuals to be interviewed; and the type or class of maturity necessary.

## 5. BENEFIT

- A Maturity Level rating assessment of quality assurance function in the perspective of auditing capability will be available

- Helps to comply with basic hygiene factor like ISO and CMMI once audit maturity level 2 is achieved
- Findings that describe the strengths and weaknesses of organisation relative to the AMM
- Consensus regarding the organization's key quality management area.
- An appraisal database in quality assurance area that the organization can continue to use to monitor quality assurance process improvement progress and to support future appraisals
- A proactive risk identification and mitigation for all projects of organisation in the area of delivery management, process, product and business area
- Engagement to execution level maturity of organization
- Align the vendors / suppliers, the organization and its customers as part of a single to reap maximum efficiencies and thus achieve business excellence

## 5. CHALLENGES

Followings are identified challenges to implement Audit Maturity Model (AMM) framework:

- The commitment from higher management (required for conducting level 5 audits) will be a key challenge as they need to understand the maturity assessment value addition based on their business objective.
- Identifying each aspect of audit checklist for each level would be crucial as this is cost effective in terms of technology, resource and training.
- The level of manual expertise at the internal or external organization level would be crucial.
- Identified findings or risks logging will be a true challenge. Coordination and further risk mitigation, in all levels, need to be synchronized to meet the business objective.

## 6. CONCLUSION

The Audit Maturity Model (AMM) and its implementation is a new concept in the area of quality assurance to unveil maturity assessment at different levels. Here a lower maturity level forms the basis of the next higher maturity level and hence it is not possible to achieve maturity of a higher level if a lower level is skipped. Hence audit maturity can be achieved stage wise from level 2 upwards. This model strengthens the organization standard process compliance at level 2 with all basic hygiene of process compliance and data quality. Level 3 focuses on process maturity and quality of deliverables by unearthing risk of product quality. At the next level, delivery outage has been prevented by proactive risk identification of delivery management area and finally, at the top level, business risks in the area of finance, customer relations, employee, infrastructure, and security. Based on the impact of business risks, varied levels of rigor are also implemented to check aspects in bottom three levels. Hence, it is a synchronized pre-emptive method of enrichment from a conventional to more business focused state. Proper mitigation of these risks can ensure success of the project and ensures customer satisfaction. The benefits identified for this framework far outweighs the challenges identified.

**Author Bhattacharya Uttam** is a Senior Consulting Manager of Cognizant Technology Solutions having 19 Years of experience in the field of strategic assessment, process definition, implementation and process improvement in CMMI, Six Sigma, and ISO 9001. Mr. Bhattacharya was born in Kolkata, India on 2nd August, 1970 and obtained his engineering graduation (Bachelor in Technology) in the year 1993 from Calcutta University, India. Mr. Bhattacharya has also completed his MBA (part time) from Calcutta University, India in 2001.

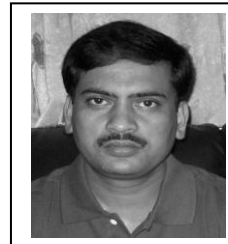

He had played the role of Quality manager for Cognizant and was responsible for ensuring quality of deliverables of the projects. He has implemented CMMI, Six Sigma, ISO 9001 framework, metrics definition for various business units in Cognizant. He has also led the CMMI assessment for Cognizant. He has wide experience in the field of consulting with direct interfacing with many clients for Strategic assessment, Process definition, implementation, improvement and maintaining their Quality Management System for the client organizations spread across geographies. He has also led a number of Six Sigma projects. He has wide experience in organization wide implementation of various processes in different types of projects and has an in-depth understanding of SDLC concepts, continual improvements and high maturity process areas.

Mr. Bhattacharya is a certified Project Management Professional (PMP®) from PMI, USA and has cleared the ITIL® version 3 Foundation Examination from Quint. He is also a certified Six Sigma Black Belt Certification form BMG, and is a certified internal auditor of ISO 9000. Mr. Bhattacharya is a certified Scrum master from Scrum Alliance and is a member of Project Management Institute (PMI), USA. He is also an eminent writer in the Cognizant Process Quality Consulting newsletter and is part of the editorial board.

**Author Rahut Amit Kumar** is a consultant of Cognizant Technology Solutions having 11 Years of experience in the field of process definition, implementation and process improvement with CMMI, Six Sigma, and ISO 9001 model. Mr. Rahut was born in Kolkata, India on 31st October, 1977 and became an engineering graduate (Bachelor in Technology) in the year 2002 from Calcutta University, India.

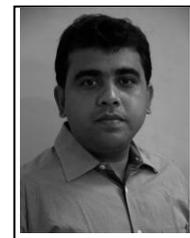

He has wide experience in the field of consulting with direct interfacing for many clients for process definition, implementation, and process improvement and maintaining their Quality Management System. He has implemented CMMI, Six Sigma, ISO 9001 framework, metrics definition for a client organization. He has worked as a Configuration Manager in the IT division of the largest private bank in Europe. He has experience in organization wide implementation of process management applications for application development and maintenance projects and has an in-depth understanding of SDLC concepts, continual improvements and high maturity process areas. He has worked as a Quality Lead for process benchmarking and implementation for a big manufacturing organization and had implemented Theory of Constraint project resulting in increased profitability.

Mr. Rahut is certified Project Management Professional (PMP®) from PMI, USA, A PRINCE2® Practitioner from APMG, UK and certified in ITIL® version 3 Foundation from APMG, UK. He is also an eminent writer in the Cognizant Process Quality Consulting newsletter and is part of the editorial board.

**Author De Sujoy** is a consultant of Cognizant Technology Solutions having 8 years of experience in various fields of Software Quality and Tool Implementation. Mr. De was born in Bankura, India on 28th of July, 1981 and received his engineering degree (Bachelor in Computer Science & Engineering) in the year 2004 from Burdwan

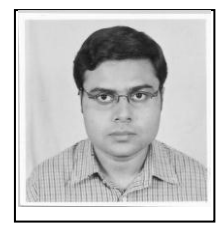

University, India, and Diploma in Business Administration in the year 2009 from Pune University, India.

He has wide experience in various fields of software quality like Process definition & implementation, process improvement and maintaining the Quality Management System. He has also experience in CMMI Level 3 implementation, ISO 9001 framework and metrics definition. He has worked as a Configuration Manager for the IT division of one of the largest private banks in Europe. He has experience in organization wide implementation of process management applications for application development and maintenance projects and has an in-depth understanding of SDLC concepts, continual improvements and high maturity process areas. In his previous organization, he was instrumental in the organization's achieving the ISO 9001:2000 recertification and its preparation for ISO 140001 certification.